\documentclass[twocolumn,smallextended]{svjour3}            
\smartqed  
\usepackage[final]{graphicx}
\usepackage{caption}
\usepackage{subfigure}
\usepackage{dcolumn}
\usepackage{bm}
\usepackage{booktabs}
\usepackage{xcolor}
\usepackage{amssymb}
\usepackage{amsmath}
\usepackage[numbers]{natbib}
\usepackage{braket}
\usepackage{float}

\newcommand{\sign}{\text{sign}}

\begin{document}

\title{Entanglement-Coherence and Discord-Coherence analytical relations for X states}


\author{J.D. Young        \and        A. Auyuanet }


\institute{J.D. Young \at
               Instituto de F\'isica, Facultad de Ingenier\'ia, Universidad de la Rep\'ublica, J. Herrera y Reissig 565, 11300, Montevideo, Uruguay \\
                        \and
           A. Auyuanet \at
              Instituto de F\'isica, Facultad de Ingenier\'ia, Universidad de la Rep\'ublica, J. Herrera y Reissig 565, 11300, Montevideo, Uruguay \\
              \email{auyuanet@fing.edu.uy}
}

\date{Received: date / Accepted: date}

\maketitle

\begin{abstract}
In this work we derive analytical relations between Entanglement and Coherence as well as between Discord and Coherence, for Bell-diagonal states and for X states, evolving under the action of several noise channels: Bit Flip, Phase Damping and Depolarizing. We demonstrate that for these families, Coherence is the fundamental correlation, that is: Coherence is necessary for the presence of Entanglement and Discord.  

\end{abstract}

\section{Introduction}
\label{intro}
Correlations can be found in quantum systems that cannot be classically described; the existence of non-classical correlations in a system is a signature that the subsystems are really quantum. The complete characterization of correlations between parts of a quantum system as well as the interrelations between these correlations is an important subject both from the fundamental and applied point of view. To understand and quantify quantum correlations is of paramount importance to comprehend the origin of the quantum advantages in quantum computing and quantum information processing. Historically, Entanglement was the first quantum correlation known \cite{EPR}, and over time was the central subject of study, both theoretically and experimentally \cite{Horodecki_RMP, HarocheEPRatoms, ZeilingerTeleportation}. Over time it was understood that there are separable states that show non-classical behavior. Quantum Discord was presented in 2001 \cite{Zurek_D} , and soon other quantum correlations related to it \cite{Modi}. Coherence, which is behind the interference phenomenon, has been a main topic in the framework of Quantum Optics. Recently Baumgratz et al. \cite{Baumgratz} performed a quantitative characterization of Coherence, where quantum coherence is treated as a physical resource. Both Entanglement and Coherence are related to the phenomenon of quantum superposition, therefore it historically was natural to try to understand qualitatively what their relationship is and if there is a quantitative relationship between them \cite{XiCoherence,Adesso_ChE,Chitambar}. That naturally extended to the study of the relation between Coherence and Discord and other related quantum correlations \cite{Vedralcoherence,Fancoherence,HU20181} .\\ 
In this work we propounded to study the existence of analytically expressible dynamic relations between Entanglement and Coherence, and between Discord and Coherence, for a simple model such as Bell-diagonal states and for a slightly more complicated one, the X states. In Sec. \ref{sec:geomBelldiagonal} we do a geometric description of the Bell-diagonal states. In Sec. \ref{sec:geometric} we present the geometric measures of correlations. In Sec. \ref{sec:Belldiag} we find analytical dynamic relations between Entanglement and Coherence and between Discord and Coherence for the Bell-diagonal states evolving under the action of several noise channels: Bit Flip, Phase Damping and Depolarizing. In Sec. \ref{sec:Xstates} we study the X states; we analize their region of existence and its variations with the parameters that describes the X states and calculate the expression of the correlations for these states. In Sec. \ref{sec:evolX} we study the evolution of the correlations and their dynamical relations under the action of the same channels applied before. In Sec. \ref{sec:Coherencefundamental} we discuss Coherence as the fundamental quantum correlation. In Sec. \ref{sec:conclu} we summarize and discuss the results.

\section{Geometry of the Bell-diagonal states}
\label{sec:geomBelldiagonal}

\begin{figure}[h]
\includegraphics[scale=0.65]{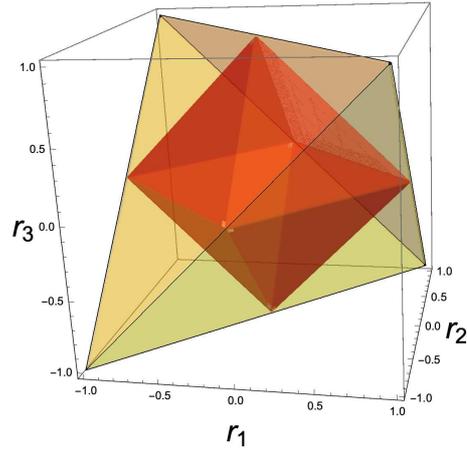}
\caption{Three-dimensional representation of the Bell-diagonal states. The tetrahedron limits the region of existence. Inside the tetrahedron, the octahedron confines the separables Bell-diagonal states.}
\label{fig:Belldiagonal0}    
\end{figure}

The Bell-diagonal states are a paradigmatic class of states in which, due to their relative mathematical simplicity, it has been possible to study the behavior of various quantum correlations.\\
They can be written as: 
\begin{equation}\label{Eq:EstadosBell}
\rho=\frac{1}{4} \Big (\mathbb{I}_2 \otimes \mathbb{I}_2 + \sum_{i=1}^{3} r_{i} \sigma_i \otimes \sigma_i \Big)
\end{equation}
where $\sigma_{i}$ are the Pauli matrices and $\vec{r}$ is the correlation vector:
\begin{equation}
\vec{r}=r_{1}\hat{i}+r_{2}\hat{j}+r_{3}\hat{k},
\end{equation}
with $r_{i}=\mathrm{Tr}(\rho\sigma_{i}\otimes\sigma_{i}),$ and 
the elements of the $\rho$ matrix can be directly related to the entries of the correlation vector as follows:
\begin{gather}
\begin{aligned}
&\rho_{11}=\rho_{44}=\frac{1}{4}(1+r_3)&\\
&\rho_{22}=\rho_{33}=\frac{1}{4}(1-r_3)&\\
&\rho_{14}=\frac{1}{4}(r_1-r_2)&\\
&\rho_{23}=\frac{1}{4}(r_1+r_2)&\\
\label{Eq:CambioDeVariable}
\end{aligned}
\end{gather}
It's well known the three-dimensional representation of the Bell-diagonal states as a function of the entries of the correlation vector \cite{HorodeckiTetrahedro}, and there are several works studying the geometrical structure of several quantum correlation in its three-dimensional representation \cite{Caves_BDS,chinosBellDiagonal,steeringbelldiagonal} . In fig.(\ref{fig:Belldiagonal0}) we can see the tetrahedron that delimits the zone of existence of the states, and the octahedron, which marks the region where the states are separable. 
Evolving under the action of different noise channels, the Bell-diagonal states describe a path within the tetrahedron \cite{Feldman17}.

\section{Geometric measures of correlations}
\label{sec:geometric}

In order to quantify quantum correlations we choose a geometric approach
\cite{GeometricEntanglement,BellomoGeometric,BellomoGeometric2,Dakic,GaussianDiscord,SquareNormDistanceCorr,geometricqc}.
 We determine how much Entanglement, Discord or Coherence a state $\rho$ posses, by its minimum distance to the set of states that doesn't posses that correlation, i.e:

\begin{eqnarray*}
E(\rho) & = & \min_{\rho^{sep}}d(\rho,\rho^{sep}),\\
D(\rho) & = & \min_{\rho^{cc}}d(\rho,\rho^{cc}),\\
C(\rho) & = & \min_{\rho^{inc}}d(\rho,\rho^{inc}),\\
\end{eqnarray*}
where $\rho^{sep}$ belongs to the set of separables states (zero Entanglement), 
$\rho^{cc}$ to the set of classical states (zero Discord) and $\rho^{inc}$. to the set of incoherent states (zero Coherence).\\
We choose to work with the Trace norm, wich is a particular case of the $p$-norm: $||A||_{p}^{p}:=\mathrm{Tr}\left(\sqrt{A^{\dagger}A}\right)^{p}$, when $p=1.$ 
The Trace norm determines $d_{1}$ as a measure of distance:
\begin{equation*}
 d_{1}(\rho,\sigma)= ||\rho-\sigma||
\end{equation*}
With the previous considerations, in the next section we will show the expressions of the three quantum correlations.

\subsection{Entanglement}
\label{sec:entanglement}
For the Entanglement we have:
\begin{equation*}
E(\rho)=\min_{\rho_{sep} \in \mathcal{S}} \frac{1}{2}||\rho-\rho_{sep}||_1=\min_{\rho_{sep} \in \mathcal{S}} Tr|\rho-\rho_{sep}|,
\end{equation*}
where $\mathcal{S}$ is the set of separable states. 
For the particular case of X states, in a previous work \cite{Feldman17} we found that: 
\begin{equation}\label{Eq:EntrelazamientoTN}
E(\rho)=2 \max \{0,|\rho_{14}|-\sqrt{\rho_{22}\rho_{33}},|\rho_{32}|-\sqrt{\rho_{11}\rho_{44}}\}
\end{equation}
which is the Concurrence of the quantum state \cite{YuEberlyConcX}.

\subsection{Discord}
\label{sec:discord}

We will use the expression of the Discord for the Bell-diagonal states developed in \cite{GDBell1,GDBell2}:

\begin{equation}
D(\rho_{BD})=\frac{r_{int}}{2},
\label{Eq:DiscordiaParaBell}
\end{equation}
where $r_{int}$ is the intermediate value of the $|r_i|$. 

\subsection{Coherence}
\label{sec:coherence}
Applying the geometric definition for the Coherence, its expression is:
\begin{equation*}
C(\rho)=\min_{\rho_{inc} \in \mathcal{I}} ||\rho-\rho_{inc}||=\sum_{\substack{i,j \\ i \ne j}} |\rho_{i,j}|.
\end{equation*}
where we used the fact that the nearest incoherent state is represented by the same $\rho$ matrix, but with all the elements out of the diagonal nulls \cite{Baumgratz}.
For the particular case of the X states, the Coherence is:
\begin{equation*}
C=|\rho_{14}|+|\rho_{32}|+|\rho_{41}|+|\rho_{23}|,
\end{equation*}
which under the assumption of all the entries of the matrix being reals, ($\rho_{14}=\rho_{41}$ y $\rho_{32}=\rho_{23}$) it has the following expression:
\begin{equation} \label{CoherenciaTN}
C=2(|\rho_{14}|+|\rho_{32}|)
\end{equation}

\section{Dynamical Relations for Bell-diagonal states}
\label{sec:Belldiag}
We will consider that our system evolves with each qubit in contact with its own environment. The evolution can be described by means of the Kraus operators \cite{nielsen2000quantum,Preskill} according to:
\begin{equation*}\label{Eq:Kraus_Locales}
\rho'_{AB}=\sum_{i,j} \big (M_i^A \otimes M_j^B  \big) \rho_{AB} \big (M_i^A \otimes M_j^B  \big)^\dagger,
\end{equation*}
where $M_i^A, M_i^B$ are the Kraus operator acting on each qubit. By means of this tool, we will study the evolution of the two qubit system under the action of several known quantum channels.

We start by writing the expression of the three correlations using the Trace Norm; the expression of the Entanglement can be obtained putting
eq.(\ref{Eq:CambioDeVariable}) in the expression of the Concurrence eq.(\ref{Eq:EntrelazamientoTN}):
\begin{equation*}
E(\rho)=\frac{1}{2} \max [0 \ , |r_1 \pm r_2| - (1\pm r_3) ]
\end{equation*}

The expression of the Discord, eq.(\ref{Eq:DiscordiaParaBell}):
\begin{equation*}\label{Eq:Discordia_TN}
D(\rho)=\textrm{int}[|r_1| \ ,|r_2| \ ,|r_3|]
\end{equation*}
And finally, for the Coherence we put eq.(\ref{Eq:CambioDeVariable}) in the expression eq.(\ref{CoherenciaTN}):
\begin{equation*}
C(\rho)=\frac{1}{2}(|r_1-r_2|+|r_1+r_2|)=\max(|r_1|,|r_2|)
\end{equation*}

In the following subsections we will study the dynamical relations between Entanglement, Discord and Coherence when the initial Bell-diagonal states evolve under the action of three known quantum channels: Bit Flip, Phase Damping and Depolarizing.

\subsection{Bit Flip}
\label{sec:bitflipBell}

The Bit flip channel models the simplest type of error that suffers a qubit: it flips the state from $\ket{0}$ to $\ket{1}$ and reciprocally, with probability $(1-p)$. 
The Kraus operators corresponding to this channel are: 
\begin{equation*}
M_1^{\textit{bf}}= \sqrt{1-p} \begin{pmatrix} 1 & 0 \\ 0 & 1 \end{pmatrix} \quad \textrm{and} \quad M_2^{\textit{bf}}=\sqrt{p} \begin{pmatrix} 0 & 1 \\ 1 & 0 \end{pmatrix}
\end{equation*}
Applying the Kraus operators to de initial Bell-diagonal state,
it's easy to see that all the information of the evolution is contained in the vector of correlations.

\begin{equation*}
\vec{r}'=r_1\hat{i}+r_2(p-1)^2\hat{j}+r_3(p-1)^2\hat{k}
\end{equation*}

The Entanglement of the evolved state is:
\begin{equation}
E(p)=\frac{1}{2} \max [0 \ , |r_1 \pm r_2(p-1)^2| - |1\pm r_3(p-1)^2| ]
\label{EntBF}
\end{equation}

The expression of the Discord:

\begin{equation}
D(p)=\textrm{int}[|r_1| \ ,|r_2|(p-1)^2 \ ,|r_3|(p-1)^2]
\label{DiscBF}
\end{equation}

And finally the Coherence:

\begin{equation}
\begin{split}
C(p)=\frac{1}{2}(|r_1-r_2(p-1)^2|+|r_1+r_2(p-1)^2|)= \\
\max (|r_1|,|r_2|(p-1)^2)
\end{split}
\label{CoBF}
\end{equation}

\subsubsection{Entanglement and Coherence}

\begin{figure}[h]
	\centering
	\includegraphics[scale=0.65]{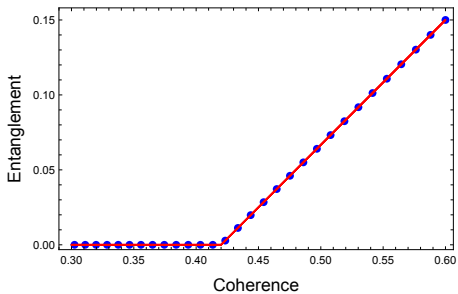}
	\caption{Entanglement as a function of the Coherence, under the action of the Bit Flip channel. $r_1=-0.3,r_2=0.6,r_3=0.4$.}
\label{fig:entcohe}
	\end{figure}

In order to express the Entanglement as a function of Coherence, we have to discriminate two alternatives: in the region of the space where $|r_1|>|r_2|(p-1)^2$,  Entanglement and Coherence are independent. On the other hand, where $|r_1|<|r_2|(p-1)^2$, we can express the Entanglement as a function of the Coherence:

\begin{equation}
E=\max \Big[0,\frac{1}{2}(|r_1 \pm  \sign{(r2)C}|- |1\pm \frac{r_3}{r_2}C|)\Big]
\label{ECBF}
\end{equation}
We show in fig.(\ref{fig:entcohe}) this relation for $r_1=-0.3$, $r_2=0.6$ and $r_3=0.4$. The blue points are the plot of Entanglement, eq.(\ref{EntBF}) versus Coherence, eq.(\ref{CoBF}), and the red line corresponds to eq.(\ref{ECBF}) Taking into account what we explained above, this equation is valid for $0< p < 0.29$. Note that in the figure the implicit parameter $p$ grows backwards.

\subsubsection{Discord and Coherence}
\begin{figure}[h]
	\centering
	\includegraphics[scale=0.65]{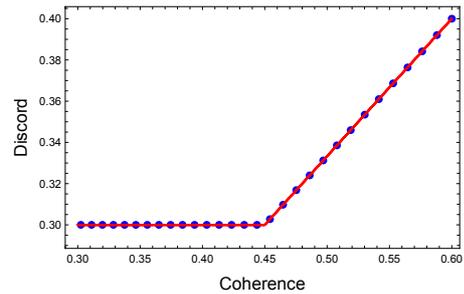}
	\caption{Relation between Discord and Coherence, under the action of the Bit Flip channel. $r_1=-0.3,r_2=0.6,r_3=0.4$. Note that for Coherence $=0.3$, $p= 0.293$;  as $p$ continues to grow, Coherence remains constant and Discord decreases quadratically regardless of Coherence. Red line and blue points explained in text.}
\label{fig:disccohe}
	\end{figure}

Looking for a functional relation between Discord and Coherence, we find that it is enough to analyze the initial values of the correlation vector. It is easy to see that this is only possible when $|r_{1}|=\min{(|r_{1}|,|r_{2}|,|r_{3}|)}$. 
Starting with $|r_1|<|r_2|<|r_3|$, as $\vec{r}$ evolves, we find two regions: one where $D=C$ (when $|r_1|<|r_2| (p-1)^2$), and another (when $|r_2| (p-1)^2<|r_1|$ )where  Discord decay quadratically, $D=|r_3|(p-1)^2$ independent of Coherence, which remains constant, $C=|r_1|$.
When initially we have $|r_1|<|r_3|<|r_2|$ we find three regions: one where $D=\frac{|r_3|}{|r_2|}C$ (when $|r_1|<|r_3| (p-1)^2$), a second region where Discord is constant, $D=|r_1|$ independent of Coherence (when $|r_3|(p-1)^2<|r_1| $), and a third region where the Discord decays quadratically, $D=|r_2|(p-1)^2$ while the Coherence remains constant, $C=|r_1|$ (when $|r_3|(p-1)^2<|r_2| (p-1)^2<|r_1| $).
We show in fig.(\ref{fig:disccohe}) one example for the case  $|r_1|<|r_3|<|r_2|$. The red line corresponds to de equation of Discord in function of Coherence; the blue points are the plot of Discord, eq.(\ref{DiscBF}) versus Coherence, eq.(\ref{CoBF}).

\subsection{Phase Damping}
\label{sec:phasedampingBell}

This channel describes a type of noise that is completely quantum: it is a process where quantum information is lost without loss of energy \cite{nielsen2000quantum}.
The Kraus operators corresponding to this channel are:
\begin{multline*}\label{Eq:Phase_Flip_Kraus}
M_0^{\textit{pd}}=\sqrt{1-p} \begin{pmatrix} 1 & 0 \\ 0 & 1 \end{pmatrix} ,\quad M_1^{\textit{pd}}=\sqrt{p}\begin{pmatrix} 1 & 0 \\ 0 & 0 \end{pmatrix} ,\\
M_2^{\textit{pd}}=\sqrt{p}\begin{pmatrix} 0 & 0 \\ 0 & 1 \end{pmatrix} 	
\end{multline*}
Applying the Kraus operators to de initial Bell-diagonal state we obtain the expression of the evolved correlation vector:
\begin{equation*}\label{Eq:R_PD}
\vec{r}'=r_1(p-1)^2\hat{i}+r_2(p-1)^2\hat{j}+r_3\hat{k},
\end{equation*}
which allow us to calculate the expressions of the three correlations under this channel.
For the Entanglement we have:
\begin{equation}
E(p)=\frac{1}{2} \max [0 \ , |r_1 \pm r_2|(p-1)^2 - |1\pm r_3| ]
\label{EntPD}
\end{equation}
The expression of the Discord is:

\begin{equation}\label{Eq:Discordia_TN_PD}
D(p)=\textrm{int}\Big[|r_1|(p-1)^2 \ ,|r_2|(p-1)^2 \ ,|r_3|\Big]
\end{equation}

The Coherence has the following expression:
\begin{equation}\label{Eq:Coherencia_TN_P}
\begin{split}
C(p)=\frac{1}{2}(|r_1-r_2|+|r_1+r_2|)(p-1)^2= \\
\max (|r_1|, \ |r_2|) (p-1)^2
\end{split}
\end{equation}

\subsubsection{Entanglement and Coherence}
For this evolution, we can always find a relation between Entanglement and Coherence:
\begin{equation}
E=\max \Big[0,\frac{C|r_1 \pm r_2|}{2\max(|r_1|,|r_2|)} - \frac{|1 \pm r_3|}{2} \Big]
\label{EnCoPD}
\end{equation}

\begin{figure}[ht]
	\centering
	\includegraphics[scale=0.65]{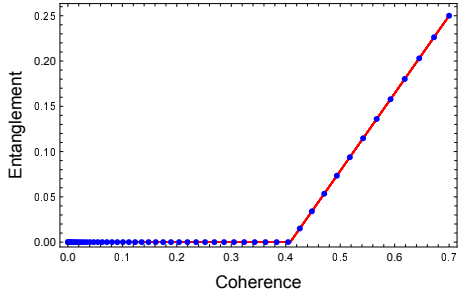}
	\caption{Entanglement as a function of Coherence under the phase damping channel.$\, r_1=-0.7,r_2=0.5$ and $r_3=0.3$. }
\label{fig:discohepd}
	\end{figure}

We show this relation in Fig.(\ref{fig:discohepd}): blue dots are the plot of Entanglement eq.(\ref{EntPD}), versus Coherence, eq. (\ref{Eq:Coherencia_TN_P}). The red line corresponds to 
eq.(\ref{EnCoPD}).

\subsubsection{Discord and Coherence}

Taking into consideration the expression for the Discord:
\begin{equation*}
	D=\textrm{int}(|r_1|(p-1)^2,|r_2|(p-1)^2,|r_3|)
\end{equation*}
we see that it is possible to relate Discord with Coherence if
$|r_1|$ or $|r_2|$ are the intermediate values. If the intermediate value is $|r_3|$, the Discord will be constant and independent of the Coherence.

When the intermediate value is $r_1$ or $r_2$ the expression of the Discord is:
\begin{equation}\label{Eq:Vinculo_CD_TNpd}
D=\frac{ |r_i| C}{\max(|r_1|,|r_2|)}, 
\end{equation}
where $r_i=r_1,r_2$, depending on which is the intermediate value. This relation is showed in fig.(\ref{fig:discohepd}), where we can distinguish clearly 3 regions: each one corresponding to $|r_1|(p-1)^2$, $|r_2|(p-1)^2$ or $|r_3|$ being the intermediate value. The red line corresponds to de equation of Discord in function of Coherence, eq.(\ref{Eq:Vinculo_CD_TNpd}); the blue points are the plot of Discord, eq.(\ref{Eq:Coherencia_TN_P}) versus Coherence, eq.(\ref{Eq:Coherencia_TN_P}).
\begin{figure}[ht]
	\centering
	\includegraphics[scale=0.65]{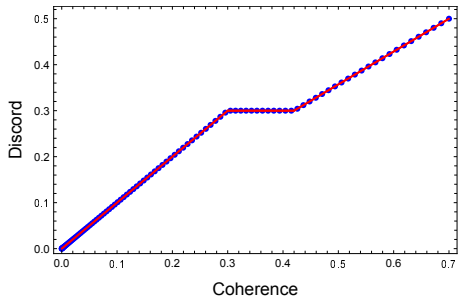}
	\caption{Discord as a function of Coherence under the phase damping channel.$\, r_1=-0.7,r_2=0.5$ and $r_3=0.3$. }
\label{fig:discohepd}
	\end{figure}

\subsection{Depolarizing}
\label{sec:depolarizingBell}

The depolarizing channel is a well-known channel; we can describe it by explaining its action on a qubit: with probability (1-p) the qubit remains unaffected and with probability p it is depolarized \cite{nielsen2000quantum}.

\begin{multline*}
M_0^d=\sqrt{1-p} \begin{pmatrix} 1 & 0 \\ 0 & 1 \end{pmatrix}, \quad M_1^d=\frac{p}{3}\begin{pmatrix} 0 & 1 \\ 1 & 0 \end{pmatrix}, \\ \qquad M_2^d=\frac{p}{3} \begin{pmatrix} 0 & -i \\ i & 0 \end{pmatrix}, \quad M_3^d=\frac{p}{3} \begin{pmatrix} 1 & 0 \\ 0 & -1 \end{pmatrix}	
\end{multline*}

The expression of the evolved correlation vector under the depolarizing channel is:
\begin{equation*}\label{Eq:R_BF}
\vec{r}'=r_1(p-1)^2+r_2(p-1)^2+r_3(p-1)^2
\end{equation*}

The Entanglement, Discord and Coherence are expressed as:

\begin{equation}
E(\rho) = \frac{1}{2} \max [0, |r_1\pm r_2|(p-1)^2 - |1\pm r_3(p-1)^2|]
\label{EntDepo}
\end{equation}
\begin{equation}
D(\rho) = \textrm{int}[|r_1| \ ,|r_2| \ ,|r_3|](p-1)^2, 
\label{DiscDepo}
\end{equation}
\begin{equation}
C(\rho) = \max(|r_1|,|r_2|)(p-1)^2. 
\label{CoheDepo}
\end{equation}
\subsubsection{Entanglement and Coherence}
Also in this case we can easily write Entanglement as a function of Coherence.
\begin{equation*}
E=\frac{1}{2}\max \Big[0,\frac{(|r_1\pm r_2|)C}{\max(|r_1|,|r_2|)} - |1\pm \frac{r_3 C}{\max(|r_1|,|r_2|)}| \Big],
\end{equation*}
and again we find a linear relation between Entanglement and Coherence.
This relation is showed in fig.(\ref{fig:entcohedep}):

\begin{figure}[ht]
	\centering
	\includegraphics[scale=0.65]{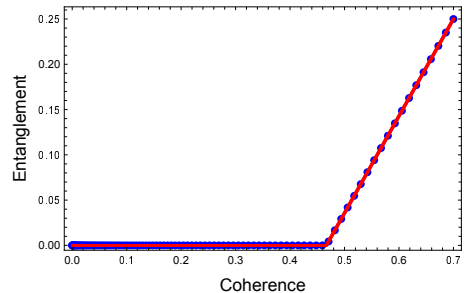}
	\caption{Entanglement as a function of Coherence under the depolarizing channel.$\, r_1=-0.7,r_2=0.5$ and $r_3=0.3$. }
\label{fig:entcohedep}
	\end{figure}

\subsubsection{Discord and Coherence}
\begin{figure}[h]
	\centering
	\includegraphics[scale=0.65]{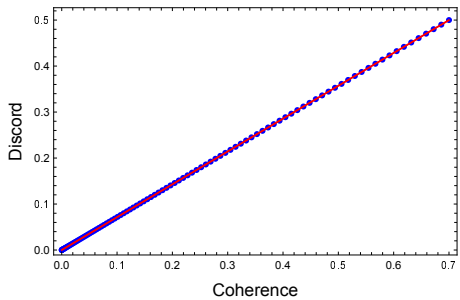}
	\caption{Discord as a function of Coherence under the depolarizing channel.$\, r_1=-0.7,r_2=0.5$ and $r_3=0.3$. }
\label{fig:discohedep}
	\end{figure}

As for the case of the previous channel, we can express Discord in function of Coherence:
\begin{equation*}\label{Eq:Vinculo_CD_TNpd}
D=\frac{C |r_i|}{\max(|r_1|,|r_2|)}, 
\end{equation*}

where $r_i=r_1,r_2,r_3$, depending which is the intermediate value. The difference with the previous channel is that for the Depolarizing channel, the intermediate value remains the same throughout the evolution because the three components of the correlations vector evolve the same way. This relation is showed in fig.(\ref{fig:discohedep}).

\section{Study of X states}
\label{sec:Xstates}

In this section we will work with the called X states, which their density matrix contains only non-zero elements along the main diagonal and anti-diagonal \cite{YuEberlyConcX}. These states, which include the Bell-diagonal states we previously studied, are described by the following expression:
\begin{equation*}
\rho = \frac{1}{4}( \mathbb{I}_4+s \cdot \sigma_3 \otimes \mathbb{I} +\mathbb{I} \otimes c \cdot \sigma_3 + \sum_{j=1}^{3} r_j  \sigma_j \otimes \sigma_j)
\end{equation*}
Unlike Bell-diagonal states that only need three parameters to be described: $(r_1,r_2,r_3)$, 
to describe the X states, it is necessary to add two additional parameters: $s$ and $c$.
In the following it will be helpful to express the elements of the density matrix $\rho_{i,j}$ as a function of the new parameters $r_i$, $s$ y $c$:  
\begin{gather*}
\begin{aligned}
&\rho_{11}=\frac{1}{4}(1+r_3+s+c) \quad \rho_{22}=\frac{1}{4}(1-r_3+s-c)&\\
&\rho_{33}=\frac{1}{4}(1-r_3-s+c) \quad \rho_{44}=\frac{1}{4}(1+r_3-s-c)&\\
&\rho_{14}=\rho_{41}=\frac{1}{4}(r_1-r_2) \quad \rho_{23}=\rho_{32}=\frac{1}{4}(r_1+r_2)&\\
\label{Eq:CambioDeVariableX}
\end{aligned}
\end{gather*}
To have a geometric 3D representation of the Bell-diagonal states is possible since they depend only on the three components of $\vec{r}$. The X states require five parameters to be described; this makes difficult to study the region of existence, since to visualize in a three-dimensional diagram it is necessary to fix at least two of them.

\subsection{Regions of existence for the X states}

\begin{figure}[h]
\begin{center}
\subfigure[$s=0.3$ and $c=0.2$]
{ \label{fig:sfig1} \includegraphics[scale=0.55]{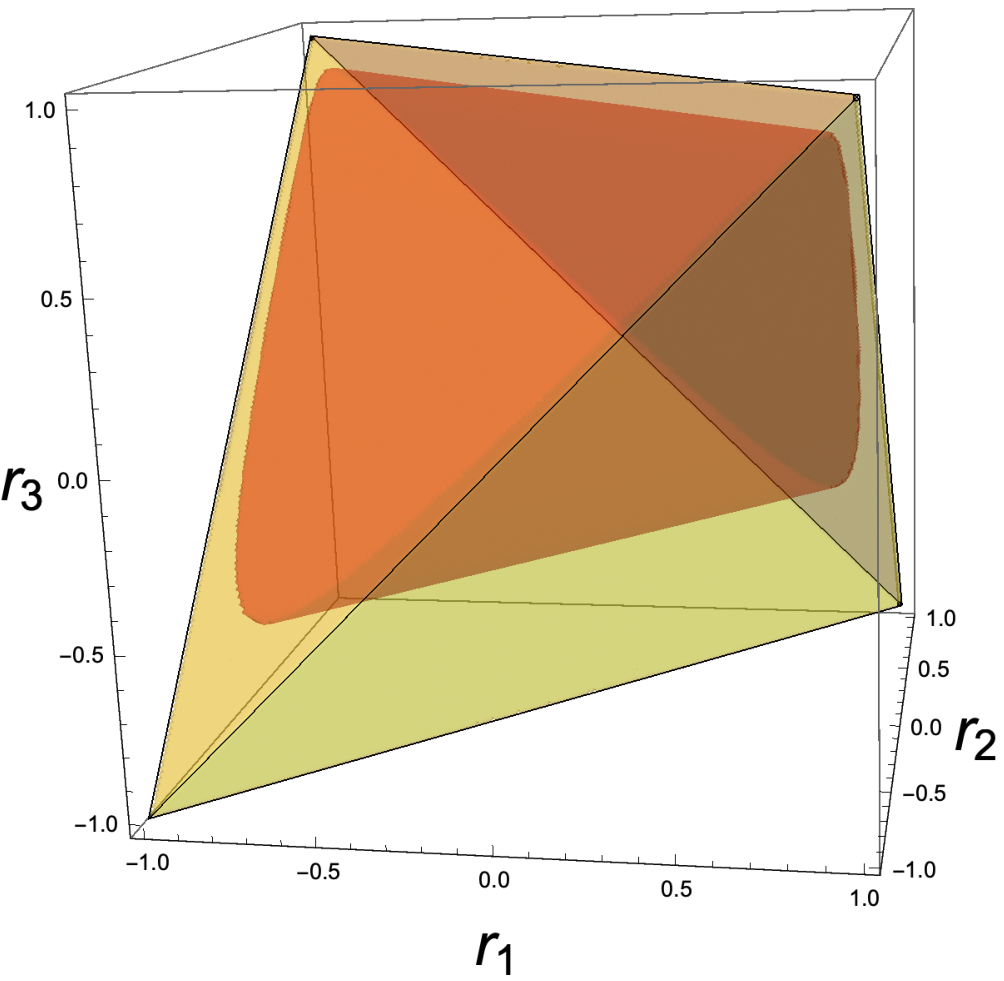}}
\subfigure[$s=0.5$ and $c=0.7$]
{ \label{fig:sfig2}  \includegraphics[scale=0.55]{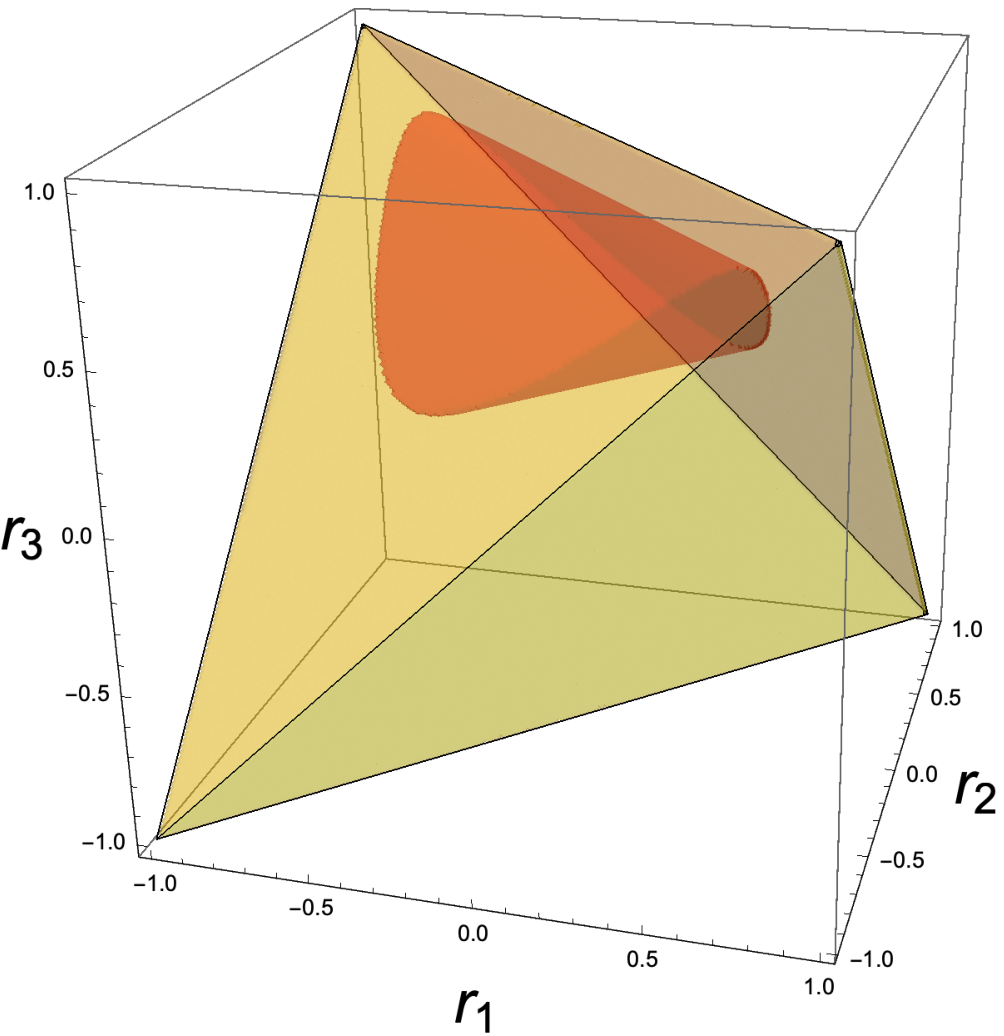}} 
\end{center}
\caption{In color red the deformed tetrahedron for different values of $s$ and $c$. The original tetrahedron was kept for comparison.}
\label{fig:tetradeformed}
\end{figure}

\begin{figure}[h]
\begin{center}
\subfigure[$s=0.3$ and $c=0.2$]
{ \label{fig:sfig3} \includegraphics[scale=0.5]{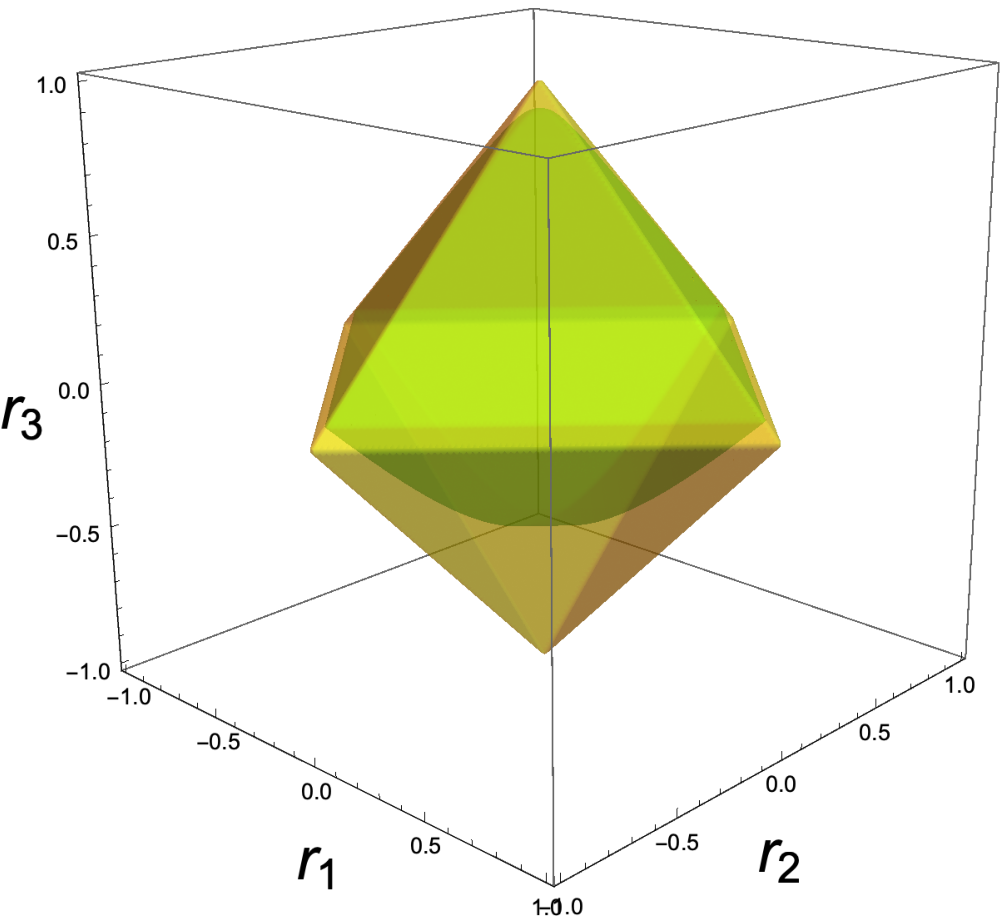}}
\subfigure[$s=0.5$ and $c=0.7$]
{ \label{fig:sfig23}  \includegraphics[scale=0.5]{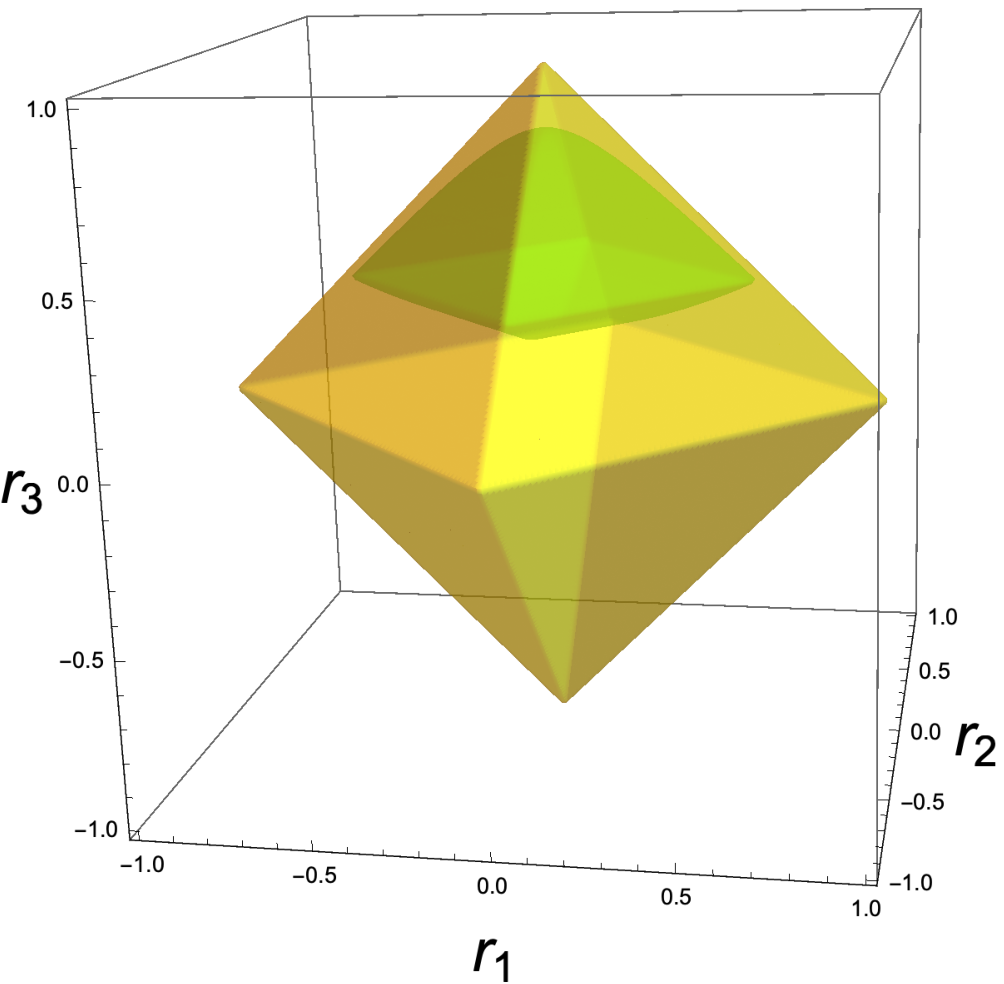}} 
\end{center}
\caption{In color green the deformed octahedron of separable states for different values of $s$ and $c$. The original octahedron was kept for comparison. }
\label{fig:tetradeformed2}
\end{figure}

In order to perform this analisis we use the positivity condition of $\rho$, which expressed in the parameters $\vec{r}$, $s$ and $c$ is:

\begin{gather*}\label{Eq:RegionExistencia}
\begin{aligned}
&|r_1-r_2| \le \sqrt{(1+r_3)^2-(s+c)^2}&\\
&|r_1+r_2| \le \sqrt{(1-r_3)^2-(s-c)^2}&
\end{aligned}
\end{gather*}
These two inequalities determine the volume of the possible X states.
By changing the values of $s$ and $c$ we can see how the region of existence is modified.
In figure (\ref{fig:tetradeformed}), we can see how the tetrahedron that determines the region of existence of the  Bell-diagonal states is deformed as the values of $s$ and $c$ change.
Just as the tetrahedron that defines the allowed states is deformed, the octahedron defined by the states of zero entanglement also undergoes deformation.
In figure (\ref{fig:tetradeformed2}), we show the transformation of the octahedron as $s$ and $c$ changes.

\subsection{Analytical expressions of correlations for X states}
We start with the Entanglement, which as we saw in eq.(\ref{Eq:EntrelazamientoTN}), for X states coincides with the expression of the Concurrence:

\begin{equation}\label{Eq:EntrelazamientoXtn}
E=\frac{1}{2} \max \big[0,|r_1 \pm r_2|-\sqrt{(1 \pm r_3)^2-(s \pm c)^2}\big]
\end{equation}

For the Discord, we will use the expression provided by \cite{Giocannetti_1ndA}:
\begin{equation}\label{Eq:DiscordiaXtn}
	D(\rho)=\begin{cases}
	\frac{|r_1|}{2}  & \text{if $\Delta >0$}.\\
	\sqrt{\frac{r_1^2 \max (r_3^2,r_2^2+s^2)-r_2^2 \min (r_3^2,r_1^2)}{\max (r_3^2,r_2^2+s^2)-\min (r_3^2,r_1^2)+r_1^2-r_2^2}} & \text{if $\Delta \leq 0$},
	\end{cases}
\end{equation}
where $\Delta=r_3^2-r_1^2-s^2$.

Finally, the expression of the Coherence is:
\begin{equation*}\label{Eq:CoherenciaXtn}
C=\max(|r_1|,|r_2|)
\end{equation*}

\subsection{Regions of Discord}
\label{ssec:regionsofDiscord}
\begin{figure}[h]
	\begin{center}
	\includegraphics[scale=0.5]{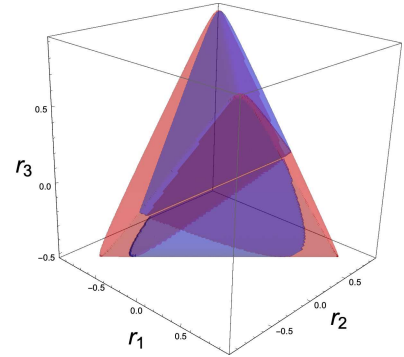}
	\end{center}
	\caption{In red, the  region of existence, in blue Region 1 with $s=0.2$.}
	\label{fig:Region1}
\end{figure}

For X states, Discord a priori, can have many different expressions in the different regions, depending on which are the maximums and minimums in eq.(\ref{Eq:DiscordiaXtn}) or which of the following relations, $r_3^2<r_1^2+s^2$ or $r_3^2 \ge r_1^2+s^2$ is verified. Taking this into consideration, we found five regions:
\begin{enumerate}
	\item $ r_3^2 > A$
	\item $\left(r_3^2 \leq A \right ) \cap \left(\max(r_3^2,B)=r_3^2\right ) \cap \left( \min(r_1^2,r_3^2)=r_3^2\right)$
	\item $\left(r_3^2 \leq A\right ) \cap \left(\max(r_3^2,B)=r_3^2\right) \cap \left(\min(r_1^2,r_3^2)=r_1^2 \right)$
	\item $\left(r_3^2 \leq A\right) \cap \left(\max(r_3^2,B)=B\right) \cap \left(\min(r_1^2,r_3^2)=r_3^2\right)$
	\item $\left(r_3^2 \leq A\right) \cap \left(\max(r_3^2,B)=B\right) \cap \left(\min(r_1^2,r_3^2)=r_1^2\right)$
\end{enumerate}
where $A=r_1^2+s^2$ and $B=r_2^2+s^2$
By substituting these results in eq. (\ref{Eq:DiscordiaXtn}) we find that in regions (1), (3) and (5) Discord has the same expression. Therefore we have to considerate three regions for Discord:
 \textbf{Region 1}, defined by:  $|r_3|>|r_1|$  where the expression of the Discord is:
\begin{equation*}
	D=\frac{|r_1|}{2},
\end{equation*}
and we show it in fig.(\ref{fig:Region1})

\begin{figure}[h]
	\begin{center}
	\includegraphics[scale=0.5]{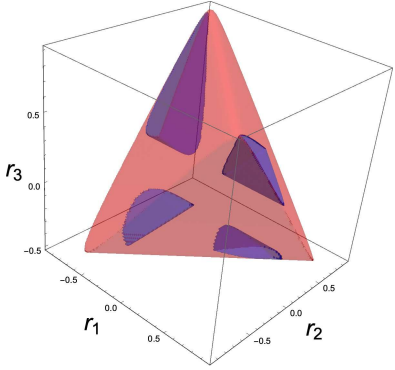}
\end{center}
	\caption{In blue Region 2, with $s=0.2$.}
	\label{fig:Region2_2}
\end{figure}

 \textbf{Region 2}, defined by:   $\left(|r_3| \leq |r_1| \right)  \cap \left(r_3^2 > r_2^2+s^2\right)$ 
 where the expression of Discord is,
\begin{equation}
	D=\frac{|r_3|}{2},
\end{equation}
and it is showed in fig. \ref{fig:Region2_2}

\begin{figure}[h]
\begin{center}
	\includegraphics[scale=0.5]{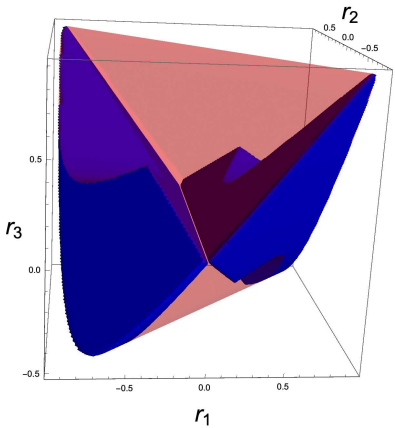}
\end{center}
	\caption{In blue Region 3, with $s=0.2$.}
	\label{fig:Region4_1}
\end{figure}

\textbf{Region 3} defined by:   $\left(|r_3| \leq |r_1| \right)  \cap \left(r_3^2 \leq r_2^2+s^2\right)$ 
 where the expression of Discord is,

\begin{equation}
D=\frac{1}{2}\sqrt{\frac{r_1^2r_2^2-r_2^2r_3^2+r_1^2s^2}{r_1^2-r_3^2+s^2}},
\end{equation}
and in fig.\ref{fig:Region4_1} we show it.

\subsection{Analysis in the $s-c$  plane}

The previous analysis was performed leaving the values of the parameters $s$ and $c$ fixed and varying $\vec{r}$. In this section we will analyze how the regions of existence change and how the correlations behave when we fix $\vec{r}$ and vary the parameters $s$ and $c$.

In the following figures \ref{fig:figENT} and \ref{fig:figDIS} we show the changes in the regions of existence and the behavior of Entanglement and Discord by varying $r_2$, on the $s-c$ plane. 
As can be seen, the region of existence contracts, until it forms a perfect square at $r_2 = 0 $, and then stretches again until it becomes a line and then disappears.
The images selected are with $ r_1 $ and $ r_3 $ fixed so that it is better understood how the region of existence is modified as $ r_2 $ is changed. Setting $ r_2 $ and $ r_3 $ fixed, and changing $ r_1 $, and fixing $ r_1 $ and $ r_2 $  and changing $ r_3 $ give similar results.

\begin{figure}[H]
\begin{center}

\subfigure[$r_1=-0.9$, $r_2=-0.08$, $r_3=0$]
{ \label{fig:sfig1} \includegraphics[scale=0.5]{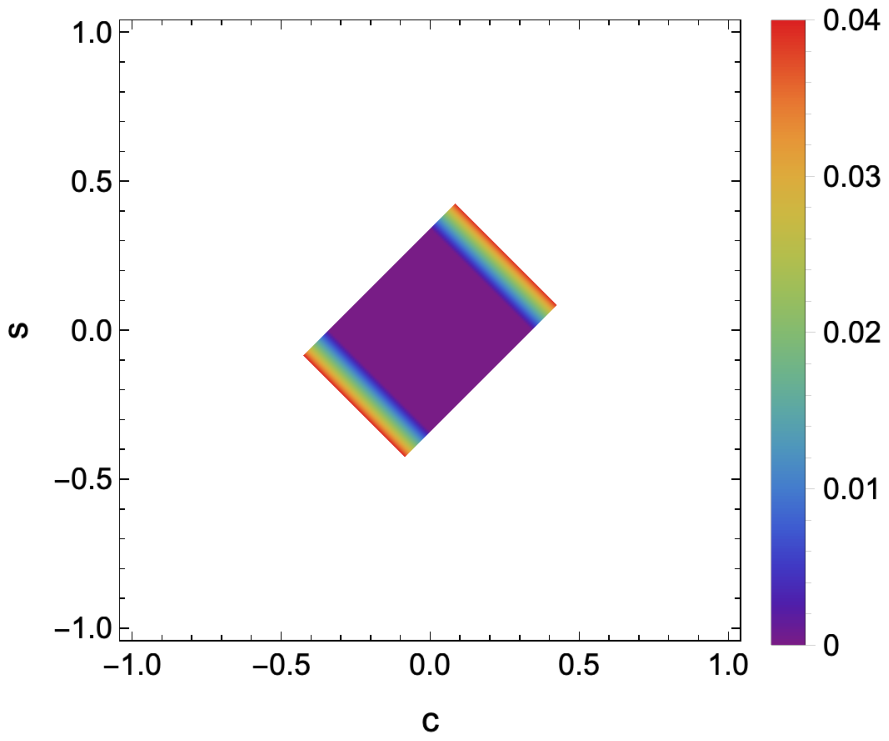}}

\subfigure[$r_1=-0.9$, $r_2=-0.04$, $r_3=0$]
{ \label{fig:sfig2}  \includegraphics[scale=0.5]{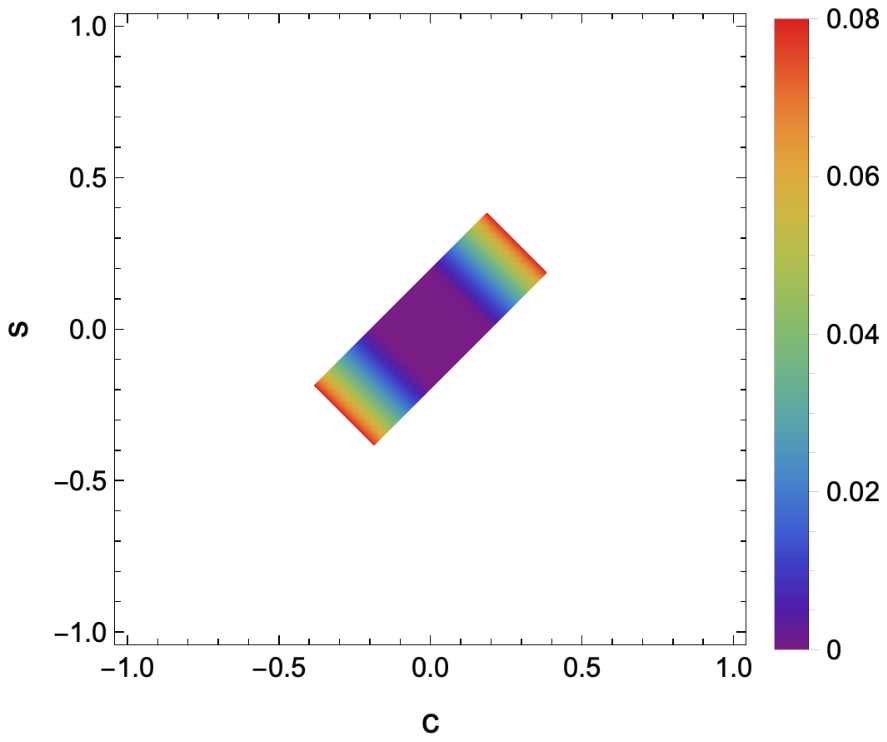}} 

\subfigure[$r_1=-0.9$ $r_2=0$ $r_3=0$]
{ \label{fig:sfig3}   \includegraphics[scale=0.5]{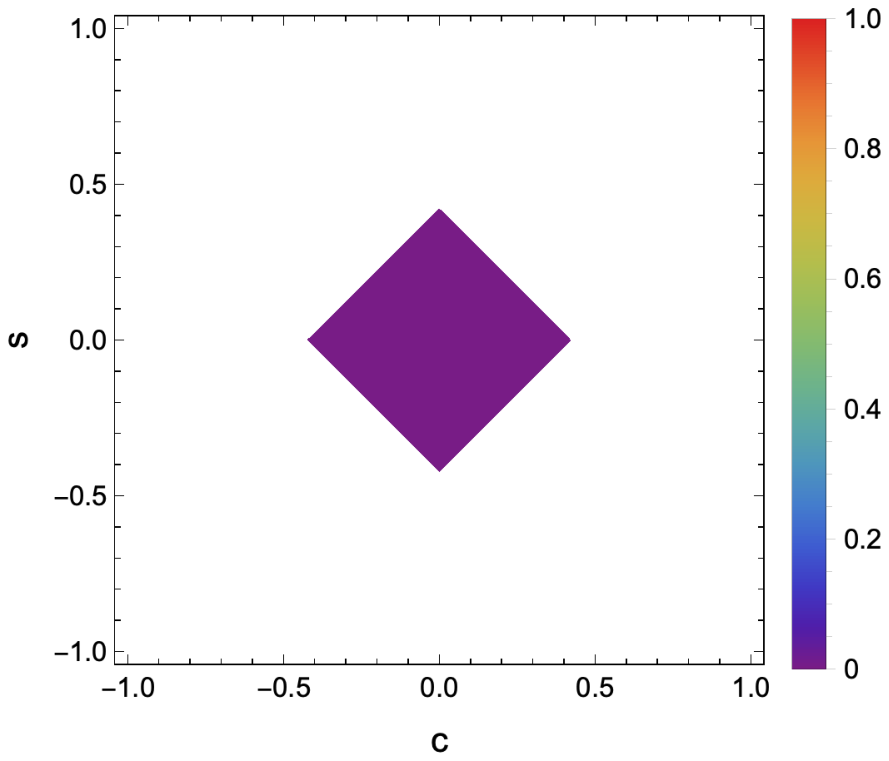}}

\subfigure[$r_1=-0.9$ $r_2=0.04$ $r_3=0$]
{\label{fig:sfig3} \includegraphics[scale=0.5]{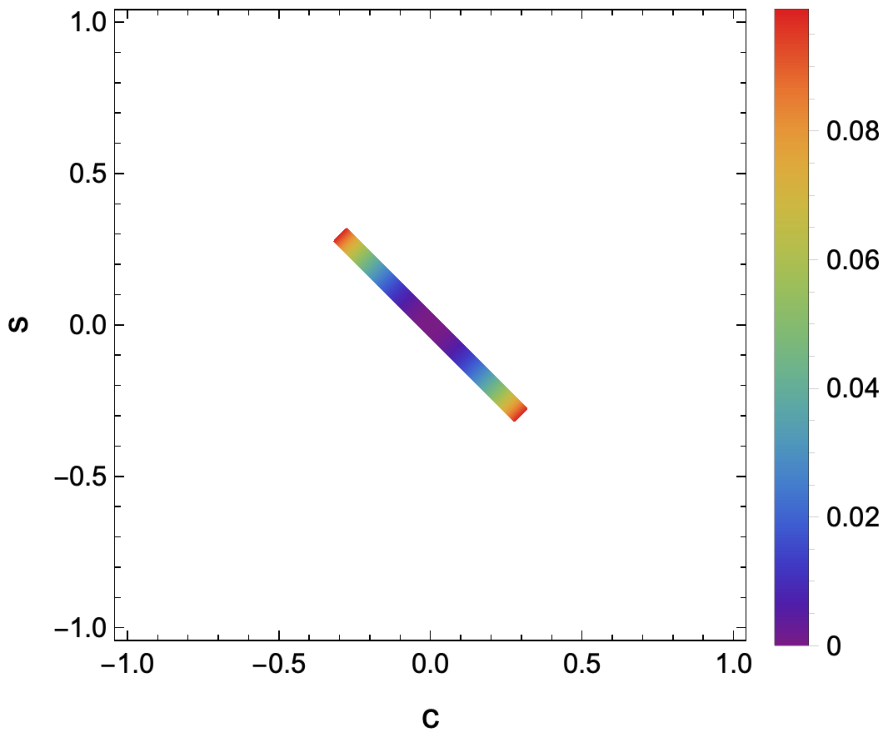}}
\end{center}

\caption{Color maps of the Entanglement in the $s-c$ plane. The white region indicates the region of non-physical states.}
\label{fig:figENT}

\end{figure}

The Discord pattern is simple, because it does not depend on parameter $c$ and grows with $s$. On the other hand, Entanglement grows as $|s \pm c|$ increase, since: $|r_1 \pm r_2|- \sqrt{(1 \pm r_3)^2 -(s\pm c)^2}$ for values of $ r_i $ within the region of existence, showing growth in the longitudinal direction.
Coherence was not included in this analysis since it is constant in this plane, because it is independent of $s$ and $c$.

\begin{figure}[H]
\begin{center}

\subfigure[$r_1=-0.9$, $r_2=-0.08$, $r_3=0$]
{ \label{fig:sfig1} \includegraphics[scale=0.5]{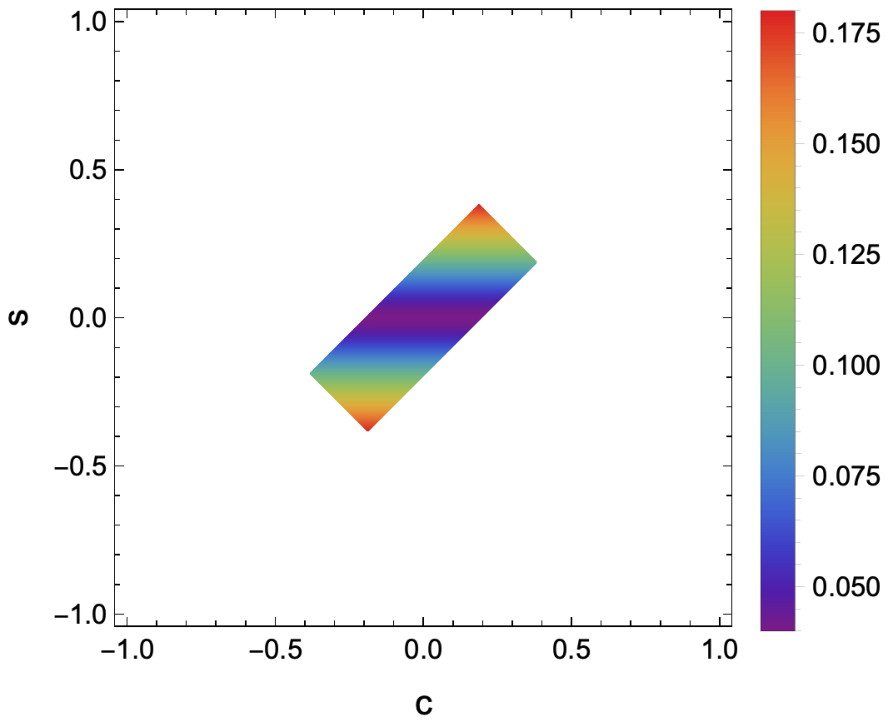}}

\subfigure[$r_1=-0.9$, $r_2=-0.04$, $r_3=0$]
{ \label{fig:sfig2}  \includegraphics[scale=0.5]{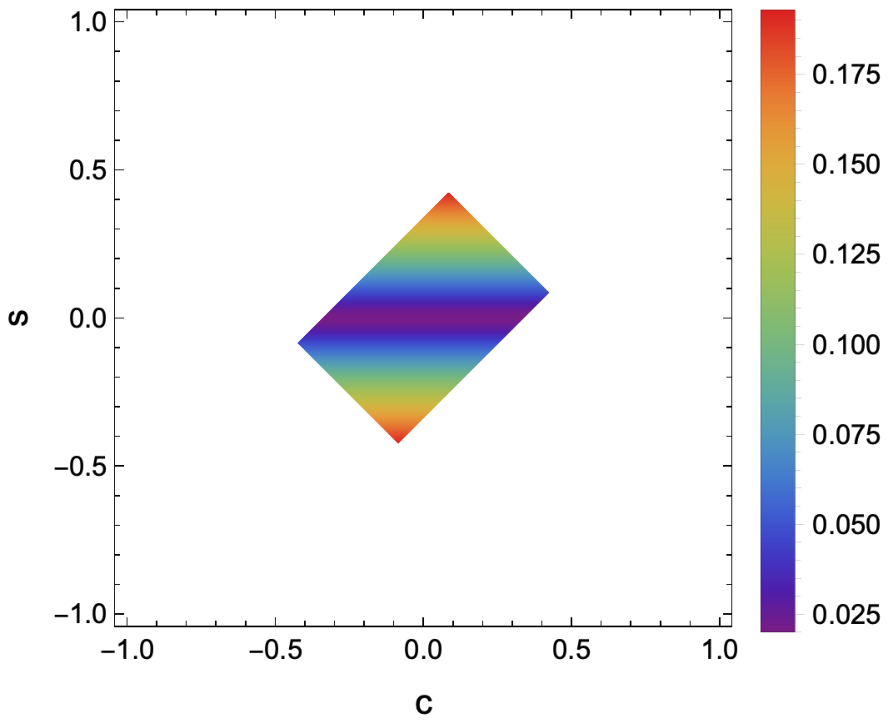}} 

\subfigure[$r_1=-0.9$ $r_2=0$ $r_3=0$]
{ \label{fig:sfig3}   \includegraphics[scale=0.5]{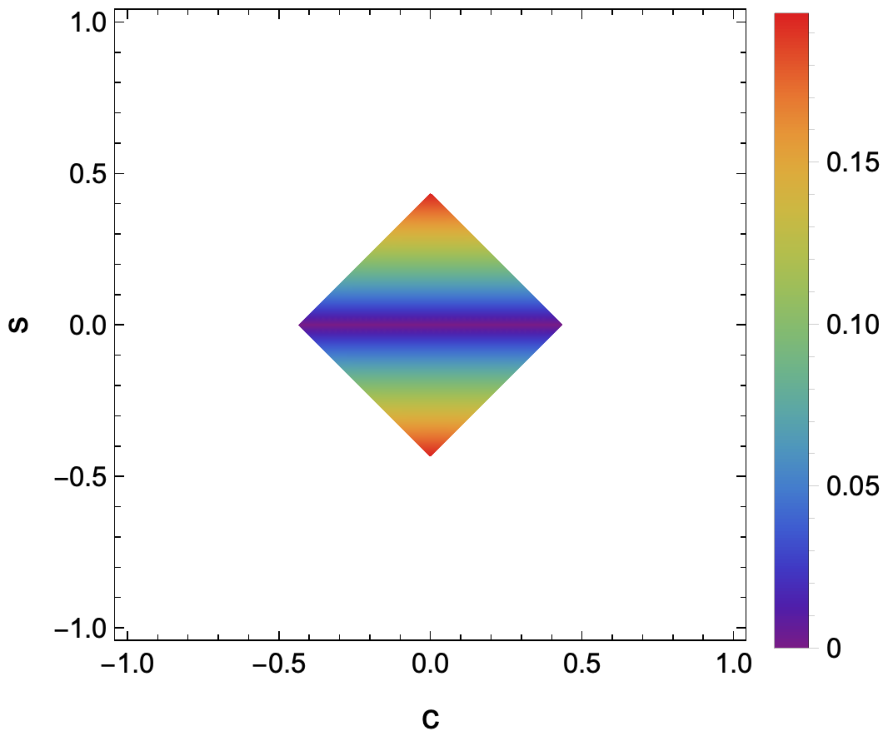}}

\subfigure[$r_1=-0.9$ $r_2=0.04$ $r_3=0$]
{\label{fig:sfig3} \includegraphics[scale=0.5]{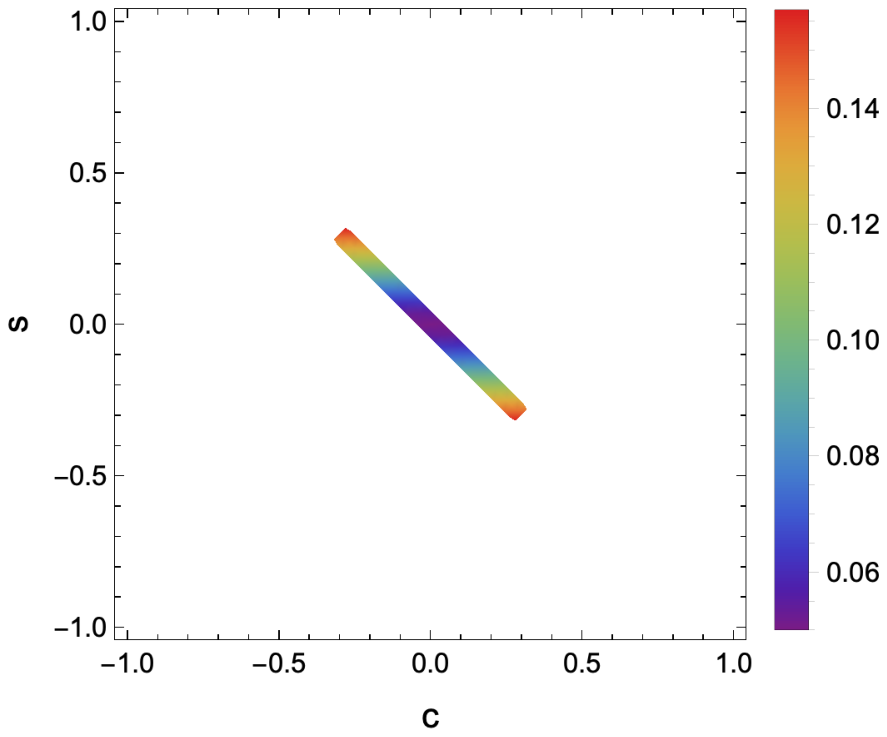}}

\end{center}

\caption{Color maps of the Discord in the $s-c$ plane. The white region indicates the region of non-physical states.}
\label{fig:figDIS}

\end{figure}

\section{Evolution of X states under noise channels}
\label{sec:evolX}

The description of the evolution of the X states is given by the evolution of the correlation vector, as in the case of the Bell-diagonal states, plus the evolution of the parameters $s$ and $c$. The only exception is made by the evolution under Phase Damping, in which the parameters $s$ and $c$ are not affected.

\subsection{Phase Damping}
\label{sec:phasedampingX}

The evolution of the correlation vector $\vec{r}$ under Phase Damping is given by:

\begin{equation*}\label{Eq:R_PD}
\vec{r}'=r_1(p-1)^2\hat{i}+r_2(p-1)^2\hat{j}+r_3\hat{k}
\end{equation*}
\subsubsection{Entanglement}
As we have already seen for the X states, each pair of values of $s$ and $c$ determines the zone of existence and the zero Entanglement octahedron. The expression of Entanglement is:
\begin{multline*}\label{Eq:EntrelazamientoXPD}
E=\frac{1}{2} \max \big[0,|r_1 \pm r_2|(p-1)^2- \\ \sqrt{(1 \pm r_3)^2 -(s \pm c)^2}\big]
\end{multline*}

\subsubsection{Coherence}
The Coherence expression of a X state that evolves under Phase Damping is simple: it decreases monotonously and reaches the value $ 0 $ when $ p = 1 $:
\begin{equation*}\label{Eq:CoherenciaXpd}
C=\max(|r_1|,|r_2|)(p-1)^2
\end{equation*}

\subsubsection{Discord}

\begin{figure}[H]
	\centering
	\includegraphics[scale=0.5]{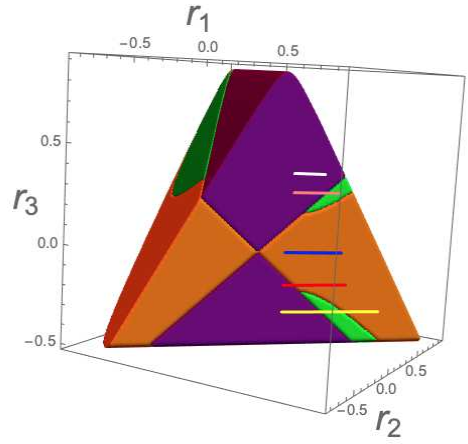}
	\caption{Cut according to the plane $r_1-r_2 = 0.2$. Region 1 is purple, Region 2 is green, and Region 3 is orange. The yellow line shows an evolution that cross through all three regions. In blue the state always remains in Region 3. In red it jumps from Region 3 to Region 1. In pink from Region 2 to Region 1. In White, it always remains within Region 1.}
	\label{Fig:EvoPDX}
\end{figure}

When a X state evolves under the Phase Damping channel, the expression of the Discord is expressed by eq.(\ref{Eq:DiscordiaXtn}):
\begin{equation*}
	D(\rho)=\begin{cases}
	\frac{|r_1|}{2}(p-1)^2  & \text{if $\Delta > 0$}.\\
	\frac{1}{2} \sqrt{\frac{A \max(r_3^2,B+s^2)-B \min(r_3^2,A)}{\max(r_3^2,B+s^2)- \min(r_3^2,A)+A-B}} & \text{if $\Delta \leq 0$},
	\end{cases}
\end{equation*}

where, $\Delta=r_3^2-s^2-A $, $A=r_1^2(p-1)^4$ and $B=r_2^2(p-1)^4.$ 
We showed from an analysis in subsection \ref{ssec:regionsofDiscord} that there are three different
regions for the Discord. Depending on the initial value of the vector $\vec{r}$ and the parameters $s$ and $c$, the state in its evolution will cross some of that regions, as shown in the figure \ref{Fig:EvoPDX}. Each color line represents a possible evolution of an X state, starting from different initial values of $\vec{r}$ and the parameters $s$ and $c$.

The figure shows how the state evolves depending on the initial conditions of $\vec{r}$. Evolution can take 5 possible paths: (1) to go through the 3 regions in descending order, Region 3, Region 2 and Region 1 (yellow path), (2) to go from Region 3 directly to Region 1 (red path) , (3) always stay in Region 3 (blue path), (4) move from Region 2 to Region 1 (pink path) or (5) always stay in Region 1 (white path). Regions 1, 2 and 3 can be written as the following inequalities: 
\begin{multline*}\label{Eq:RegionesDiscordPD}
Region 1: |r_3|>|r_1|(p-1)^2 \\
Region 2: \left(|r_3|\leq|r_1|(p-1)^2\right)  \cap\ \left(r_3^2 > r_2^2(p-1)^4+s^2\right)   \\
Region 3:  \left(|r_3|\leq|r_1|(p-1)^2\right)  \cap  \left( r_3^2 \leq r_2^2(p-1)^4+s^2\right) 
\end{multline*}
Therefore the quantum state will remain in Region 1 as long as it is verified that:
\begin{equation*}\label{Eq:Region1p}
	p>1-\sqrt{\frac{|r_3|}{|r_1|}}=p_{1},
\end{equation*}
and the expression of the Discord in this region is:
\begin{equation}
D=\frac{|r_1|}{2}(p-1)^2
\end{equation}
Note that if $r_1=0$ the state always stay in Region 1. \\
The quantum state will remain in Region 2 as long as it is verified that:
\begin{equation}\label{Eq:condicionesR2}
	\begin{cases}
	p \leq1-\sqrt{\frac{|r_3|}{|r_1|}}=p_1\\p>1-\sqrt[4]{\frac{r_3^2-s^2}{r_2^2}}=p_2,
	\end{cases}
\end{equation}
and the expression of the Discord in this region is: 
\begin{equation*}
D=\frac{|r_3|}{2}.
\end{equation*}
Note that if $|r_3|<|s|$, the state cannot be in Region 2.\\
Finally, it will remain in Region 3 as long as:
\begin{equation}
\begin{cases}
p \leq 1-\sqrt{\frac{|r_3|}{|r_1|}}=p_1\\ p<1-\sqrt[4]{\frac{r_3^2-s^2}{r_2^2}}=p_2\\
\end{cases}
\end{equation}
The expression of the Discord in this region is:
\begin{equation*}\label{Eq:DiscordiaPDX}
D=\frac{(p-1)^2}{2}\sqrt{\frac{r_1^2r_2^2(p-1)^4-r_2^2r_3^2+r_1^2s^2}{r_1^2(p-1)^4-r_3^2+s^2}}
\end{equation*}
If $p<p_1$ and $r_2=0$, the state will be in Region 2 or Region 3 depending on $|r_3|>|s|$ or $|r_3|\leq |s|$ respectively.

Looking at the figure, it is clear that the state in its evolution not necessarily crosses the three regions. We will clarify this point by giving some examples.
Let's consider an initial state defined by the following values of its parameters: $r_1=-0.6, r_2=0.4, r_3=0.3, s=0.2, c=0.3$. For these values, the state is in Region 3 and
$p_1=0.29$, and $p_2=0.25$, so $p_1>p_2$. This imply that the quantum state: remains in Region 3 when $p<p_2$,  it is in Region 2 when $p_2<p \leq p_1$ and it is in the Region 1 for $p>p_1$. This behavior is showed in the yellow path in fig.(\ref{Fig:EvoPDX}).
 Let's consider a second example with initial values $ r_1 = 0.5, r_2 = -0.2, r_3 = 0.3, s = 0.2, c = 0.3 $. In this case $ p_1 = 0.33 $ and $ p_2 = -0.06. $ $ p_2 $ is excluded since $ p $ can never take negative values. This will imply that the state will never be in Region 3, since $ p> p_2  \forall p $. The state will be in Region 2 for $ 0 <p \leq p_1 $ and in Region 1 for $ p> p_1 $. This is illustrated in the red path.
As a last example, consider the following initial conditions: $ r_1 = -0.6, r_2 = 0.4, r_3 = 0.7, s = 0.2, c = 0.3 $. In this situation the state initially is in Region 1 and $ p_1=-0.08$ , so $ p> p_1\, \forall p $ and the state always remains in Region 1, as can be seen in the white path of the figure \ref{Fig:EvoPDX}.

\subsubsection{Entanglement and Coherence}
In the region outside the octahedron of the separable states, the Entanglement and the Coherence can be related by a linear function:

\begin{equation*}\label{Eq:EntrelazCoherenciaXPD}
E=\frac{1}{2} \max \big[0,\frac{|r_1 \pm r_2| C}{\max(|r_1|,|r_2|)}- \sqrt{(1 \pm r_3)^2-(s\pm c)^2}\big]
\end{equation*}
\subsubsection{Discord and Coherence}
In order to determine the relation between Discord and Coherence, the behavior of the Discord must be taken into account in each of the three regions defined above.
In Region 1, the relation is linear:
\begin{equation*}
D=\frac{|r_1|C}{2\max(|r_1|,|r_2|)}
\end{equation*}
In Region 2, there is no relation between Discord and Coherence because the Discord is constant and depends only of $r_3$: $D=\frac{|r_3|}{2}$\\
In Region 3 the relation is:
\begin{equation*}\label{Eq:DiscordCoherencePDX}
D=\frac{1}{2}\sqrt{\frac{C^2 \left(\frac{r_1^2r_2^2C^2}{\max(|r_1|,|r_2|)^2}-r_2^2r_3^2+r_1^2s^2\right)}{r_1^2 C^2-r_3^2+s^2}}
\end{equation*}

\subsection{Bit Flip}
\label{sec:bitflipX}
As the state evolves under this channel, we see that not only does $\vec{r}$ change, but also parameters $s$ and $c$, as follows:
\begin{gather*}
\begin{aligned}
&\vec{r}'=r_1\hat{i}+r_2(p-1)^2\hat{j}+r_3(p-1)^2\hat{k} \\
&s'=s(p-1) \quad c'=c(p-1),
\end{aligned}
\end{gather*}
and this allows us to find the expression of the different correlations. \\
The Entanglement has the following expression:
\begin{equation*}
\begin{split}
E & =\frac{1}{2} \max \Big[0,|r_1 \pm r_2(p-1)^2| \\
& -\sqrt{(1 \pm r_3(p-1)^2)^2-(s \pm c)^2(p-1)^2}\Big]
\end{split}
\end{equation*}
The expression of the Coherence is:
\begin{equation*}
C=\max(|r_1|,|r_2|(p-1)^2)
\end{equation*}
\subsubsection{Discord}
According to eq.(\ref{Eq:DiscordiaXtn}) the expression for the Discord for a X state evolving under the Bit Flip channel is:
\begin{equation*}
	D(\rho)=\begin{cases}
	\frac{|r_1|}{2}    & \text{if $\Delta >0$}.\\
	\frac{1}{2} \sqrt{\frac{r_1^2 \max(A,C)-B \min(A,r_1^2)}{\max(A,C)-\min(A,r_1^2)+r_1^2-B}} & \text{if $\Delta \leq 0$},
	\end{cases}
\end{equation*}
where $A=r_3^2(p-1)^4$, $B=r_2^2(p-1)^4$, $C=r_2^2(p-1)^4+s^2(p-1)^2$ and $\Delta=A-r_1^2-s^2(p-1)^2.$

Again, to study Discord we have to analyze its behavior in 3 different regions, defined by the following relations:
\begin{multline*}\label{Eq:RegionesDiscordBF}
Region 1: |r_3|(p-1)^2>|r_1| \\
Region 2: \left(|r_3|(p-1)^2 \leq |r_1|\right)  \cap \left( (r_3^2- r_2^2)(p-1)^2 > s^2\right) \\
Region 3:  \left(|r_3|(p-1)^2 \leq |r_1|\right)   \cap \left( (r_3^2- r_2^2)(p-1)^2 \leq s^2\right) 
\end{multline*}
Carrying out the same analysis as for the Phase Damping case, we arrived at the 
following results: the X state will be in Region 1 provided that:
\begin{equation*}
\begin{cases}
p<1-\sqrt{\frac{|r_1|}{|r_3|}}, \\
r_3 \neq 0
\end{cases}
\end{equation*}
In this region the expression of Discord is:
\begin{equation*}
D=\frac{|r_1|}{2}
\end{equation*}
The state will remain in Region 2 as long as it is verified that
\begin{equation*}
\begin{cases}
p \ge 1-\sqrt{\frac{|r_1|}{|r_3|}} \\ p<1-\sqrt{\frac{s^2}{r_3^2-r_2^2}} 
\end{cases}
\end{equation*}
The expression of Discord in this region is:
\begin{equation*}
D=\frac{|r_3|}{2}(p-1)^2
\end{equation*}
Finally Region 3 it's defined by:
\begin{equation*}
\begin{cases}
p \ge 1-\sqrt{\frac{|r_1|}{|r_3|}} \\p>1-\sqrt{\frac{s^2}{r_3^2-r_2^2}}
\end{cases}
\end{equation*}
Note that if $r_3^2 \leq r_2^2$, or $r_3=0$, the state is in Region 3.
In this region the Discord is expressed as:
\begin{equation*}\label{}
D=\frac{(p-1)}{2}\sqrt{\frac{r_1^2r_2^2(p-1)^2-r_2^2r_3^2(p-1)^6+r_1^2s^2}{r_1^2-r_3^2(p-1)^4+s^2(p-1)^2}}
\end{equation*}
It should be noted that since the Bit Flip channel also changes the values of $s$ and $c$, it is not possible to show an evolution crossing the different regions, since for each value of s and c the entire volume that defines the space of X states as well as the interior regions varies.
\subsubsection{Entanglement and Coherence}
In order to relate the Entanglement and the Coherence we must distinguish between two regions outside the region of separable states:
Region A: $|r_1|>|r_2|(p-1)^2 $, in this region the region the Coherence is constant: $C=|r_1|$, and
Region B:
 $|r_1|<|r_2|(p-1)^2$, where 
 we can relate Entanglement and Coherence as follows:
\begin{equation*}
E=\frac{1}{2}\max \Bigg[0,|r_1 \pm C|-\sqrt{\Big(1\pm \frac{r_3 C}{|r_2|}\Big)^2-(s \pm c)^2\frac{C}{|r_2|}}\, \Bigg]
\end{equation*}

\subsubsection{Discord and Coherence}
By seeking to relate Discord to Coherence we have to take into account the 3 regions of Discord. In Region 1, defined by $|r_3|(p-1)^2>|r_1|$, Discord has a constant value, $D=\frac{|r_1|}{2}$ independent of the Coherence.\\
In Region 2, $|r_2|(p-1)^2 < |r_1|$ is verified, so the Coherence takes a constant value $C=|r_1|$ and Discord has the following expression: $D=\frac{|r_3|}{2}(p-1)^2$; there is no relation between them in this region neither.\\
Within Region 3, in the area where $|r_2|(p-1)^2<|r_1|$, the Coherence is constant $C=|r_1|$. \\
However, in the zone in which $|r_2|(p-1)^2>|r_1|$, Coherence takes the following expression: $C=|r_2|(p-1)^2$ and therefore we can relate it to Discord, as follows:
\begin{equation*}\label{Eq:DiscordCoherenceBFX}
D=\frac{1}{2}\sqrt{\frac{C \left(r_1^2|r_2|s+r_1^2|r_2|^2C-r_3^2C^3\right)}{r_1^2|r_2|^2+s^2|r_3|C-r_3^2C^2}}
\end{equation*}

\subsection{Depolarizing}
\label{sec:depolarizingX}

By evolving under de Depolarizing channel the correlations vector, $\vec{r}$, $s$ and $c$ change as:
\begin{gather*} \label{C=0}
\begin{aligned}
&\vec{r}'=r_1(p-1)^2\hat{i}+r_2(p-1)^2\hat{j}+r_3(p-1)^2\hat{k}&\\
&s'=s(p-1) \quad c'=c(p-1)&
\end{aligned}
\end{gather*}

The expression of the Entanglement of the X state under this channel is:
\begin{equation*}
\begin{split}
E&=\frac{1}{2} \max \Bigg[0,|r_1 \pm r_2|(p-1)^2-\\&\sqrt{\Big(1 \pm r_3(p-1)^2\Big)^2-(p-1)^2(s \pm c)^2}\,\,\Bigg]
\end{split}
\end{equation*}
The Coherence is:
\begin{equation*}
C=\max(|r_1|,|r_2|)(p-1)^2
\end{equation*}

\subsubsection{Regions of Discord}

The general expression for the Discord under this channel is:
\begin{equation*}
	D(\rho)=\begin{cases}
	\frac{|r_1|}{2}(p-1)^2\\
	\frac{(p-1)^2}{2} \sqrt{\frac{r_1^2 A-r_2^2(p-1)^2 \min(r_3^2,r_1^2)}{A-(p-1)^2( r_1^2-r_2^2-\min(r_3^2,r_1^2))}}
	\end{cases}
\end{equation*}
where $A= \max(r_3^2(p-1)^2,r_2^2(p-1)^2+s^2)$ The upper expression corresponds to the condition $r_1^2(p-1)^2-r_3^2(p-1)^2+s^2< 0$ and the lower to $r_1^2(p-1)^2-r_3^2(p-1)^2+s^2 \ge 0. $
This again offers three regions with different expression for Discord. After a little mathematical manipulation they are:

\begin{multline*}\label{Eq:RegionesDiscordDepo}
Region 1: |r_3|>|r_1| \\
Region 2:  |r_3|\leq|r_1|  \cap \left(r_3^2(p-1)^2 > r_2^2(p-1)^2+s^2  \right)  \\
Region 3:|r_3|\leq|r_1| \cap \left(r_3^2(p-1)^2 \leq r_2^2(p-1)^2+s^2 \right)  
\end{multline*}
In Region 1, the Discord is expressed as:
\begin{equation*}
D=\frac{|r_1|}{2}(p-1)^2
\end{equation*}
The state will remain in Region 2 as long as it is verified that:

\begin{equation*}
\begin{cases}
|r_3|<|r_1|\\p<1-\sqrt{\frac{s^2}{r_3^2-r_2^2}}\\
\end{cases}
\end{equation*}
and the Discord in this region is expressed as:
\begin{equation*}
D=\frac{|r_3|}{2}(p-1)^2.
\end{equation*}
If $|r_3|\leq |r_2|$ the state will be in Region 3. \\
The quantum X state will remain in Region 3 as long as:
\begin{equation*}
\begin{cases}
|r_3|<|r_1|\\p>1-\sqrt{\frac{s^2}{r_3^2-r_2^2}}\\
\end{cases}
\end{equation*}
and in this region Discord is expressed as:
\begin{equation*}\label{}
D=\frac{(p-1)}{2}\sqrt{\frac{r_2^2(p-1)^2(r_1^2-r_3^2)+r_1^2s^2}{2r_2^2-r_1^2+r_3^2}}
\end{equation*}

\subsubsection{Entanglement and Coherence}
It is also possible to relate in this case, the Entanglement and the Coherence:
\begin{multline*}
E=\frac{1}{2}\max\Bigg[0,|r_1\pm r_2|\frac{C}{\max(|r_1|,|r_2|)}-\\ \sqrt{\Bigg(1\pm \frac{Cr_3}{\max(|r_1|,|r_2|)}\Bigg)^2-\frac{C(s \pm c)^2}{\max(|r_1|,|r_2|)} }\,\,\Bigg]
\end{multline*}

\subsubsection{Discord and Coherence}
To link Discord and Coherence we must analyze the behavior of both magnitudes in each of the regions that we already discussed. It is immediate to show that in regions 1 and 2 the relation is linear.
In Region 1:
\begin{equation*}
D=\frac{|r_1|C}{2\max(|r_1|,|r_2|)}
\end{equation*}
In Region 2:
\begin{equation*}
D=\frac{|r_3|C}{2\max(|r_1|,|r_2|)}
\end{equation*}
In Region 3, the relation is a little bit complicated:
\begin{equation*}
D=\frac{1}{2}\sqrt{\frac{C (r_2^2(r_1^2-r_3^2)C+\max(|r_1|,|r_2|)r_1^2s^2)}{ (\max(|r_1|,|r_2|)^2)(2r_2^2-r_1^2+r_3^2)}}
\end{equation*}

\section{Coherence as the fundamental correlation}
\label{sec:Coherencefundamental}

As the theories of quantum correlations have been developed, attempts have been made to understand the possible relationships between them \cite{Bruss_ED_M,Vedralcoherence,XiCoherence,Fancoherence}.
A reasonable question to ask is whether there is a correlation that is fundamental, that is, if there is a correlation whose presence is necessary for the others to exist.
We begin by analyzing the expression of Coherence:
\begin{equation*}
C=\max(|r_1|,|r_2|),
\end{equation*}
We note that for Coherence to be canceled, it is necessary and sufficient that $r_1=r_2=0.$
If we impose that condition in the expression of the Entanglement:
\begin{equation*}
E=\frac{1}{2} \max \big[0,|r_1 \pm r_2|-\sqrt{(1 \pm r_3)^2-(s \pm c)^2}\big],
\end{equation*}
we observe that necessarily the Entanglement is zero. Therefore we can affirm that: 
\begin{equation*}
C=0 \Rightarrow E=0.
\end{equation*}
Making the same analysis with the expressions of Discord eq.(\ref{Eq:DiscordiaXtn}):
\begin{equation}
D=\frac{|r_1|}{2}  \iff r_3^2>r_1^2+s^2
\end{equation}
using that $r_1=0$, we obtain,
\begin{equation*}
D=0  \iff |r_3|>|s|
\end{equation*}
For the complementary region, $r_3^2<r_1^2+s^2$:
\begin{equation*}
D=\frac{1}{2}\sqrt{\frac{r_1^2 \max (r_3^2,r_2^2+s^2)-r_2^2 \min (r_3^2,r_1^2)}{\max (r_3^2,r_2^2+s^2)-\min (r_3^2,r_1^2)+r_1^2-r_2^2}},
\end{equation*}
imposing again $r_1=r_2=0$, we obtain:
\begin{equation*}
D=0  \iff |r_3|<|s|
\end{equation*}
From these results, we can conclude that:
\begin{gather*} \label{C=0}
\begin{aligned}
&C=0 \Rightarrow E=0&\\
&C=0 \Rightarrow D=0&
\end{aligned}
\end{gather*}
What if the Entanglement is zero? What consequences does this have on Discord and Coherence? By canceling the Entanglement, in expression eq.(\ref{Eq:EntrelazamientoXtn}), we obtain the region of separable states, where we have already seen that there are states with Coherence and non-null Discord. \\
Now we will analyze the consequences of nullifying Discord in the other correlations. We should consider the two regions of Discord.
In the region $r_3^2>r_1^2+s^2$, $D=\frac{|r_1|}{2}$. Whe then have, that:
\begin{equation*}
D=0 \iff r_1=0.
\end{equation*}
Clearly for the states in which it is verified that $|r_2|\neq 0$, the Discord being zero does not imply that the Coherence is zero.\\
In the region $r_3^2>r_1^2+s^2$, the expression of the Discord is:
\begin{equation*}
D=\frac{1}{2}\sqrt{\frac{r_1^2 \max (r_3^2,r_2^2+s^2)-r_2^2 \min (r_3^2,r_1^2)}{\max (r_3^2,r_2^2+s^2)-\min (r_3^2,r_1^2)+r_1^2-r_2^2}}
\end{equation*}
The condition for the Discord to be zero is:
\begin{equation}\label{Eq:CondicionesDiscordia}
D=0 \iff r_1^2\max(r_3^2,r_2^2+s^2)=r_2^2\min(r_3^2,r_1^2)
\end{equation}
Here we must consider four cases:
\begin{enumerate}
	\item $\max(r_3^2,r_2^2+s^2)=r_3^2$ and $\min(r_3^2,r_1^2)=r_3^2$
	\item $\max(r_3^2,r_2^2+s^2)=r_2^2+s^2$ and $\min(r_3^2,r_1^2)=r_3^2$
	\item $\max(r_3^2,r_2^2+s^2)=r_3^2$ and $\min(r_3^2,r_1^2)=r_1^2$
	\item $\max(r_3^2,r_2^2+s^2)=r_2^2+s^2$ and $\min(r_3^2,r_1^2)=r_1^2$
\end{enumerate}
By substituing each one in eq.(\ref{Eq:CondicionesDiscordia}) we obtain:
\begin{enumerate}
	\item $r_1=r_2$
	\item $r_3^2=\frac{r_1^2}{r_2^2}(r_2^2+s^2)$
	\item $r_1=r_2$
	\item $s=0$
\end{enumerate}
Neither of these conditions implies zero Coherence or zero Entanglement.
This result is very important and shows that for the X states, Coherence is the fundamental correlation.\\
All this results point in the same direction than Stretslov \cite{Adesso_ChE} , who showed that from a coherent state it is possible to produce Entanglement, but that it is impossible to produce it from an incoherent state. 

\section{Conclusions}
\label{sec:conclu}
In this work we derive analytical relations between Entanglement and Coherence as well as between Discord and Coherence, for Bell-diagonal states and for the X states, evolving under the action of several noise channels: Bit Flip, Phase Damping and Depolarizing. For the Bell-diagonal states  we found analytical relations between Entanglement and Coherence, and Discord and Coherence, for all cases. In particular all relations between these correlations are linear.
The study of the X states has the additional complexity that 5 parameters are required to describe the family, which modifies the region of existence: the tetrahedron of the Bell-diagonal states compresses and deforms as the two new parameters $s$ and $c$ change.
For the X states we described the different regions that determine the expression of Discord and in each of these regions we discused the existence of an analytical relation with Coherence. 
We found that Discord defines three different regions according to its analytical expression: there are regions where Discord remains constant regardless of Coherence, regions where it decays quadratically without being able to relate to it, and regions where it depends on Coherence in a more or less trivial way depending on the channel applied.
Finally we demonstrate that for the families of studied states, Coherence is the fundamental correlation, that is: Coherence is necessary for the presence of Entanglement and Discord.   
Our result points in the same line as previous results: Kok Chuan Tan et. al demonstrated that the Correlated Coherence in a bipartite system is necessary for the system to have Discord \cite{Tan}, and Streltsov et. al showed that Coherence is necessary to generate Entanglement
 \cite{Adesso_ChE}.

\begin{acknowledgements}
J.D. Young acknowledges the scholarship of Agencia Nacional de Investigaci\'{o}n y Formaci\'{o}n (ANII).
\end{acknowledgements}

%
%



%
%

\end{document}